\documentclass[12pt]{article}

\input epsf
\def\beq{\begin{eqnarray}}
\def\eeq{\end{eqnarray}}
\def\m{M_*}
\def\mpl{M_{\rm Pl}}

\def\E{{\cal E}}
\def\bo{{\nabla^2}}

\def\p{8\,\pi\,G_N}

\def\lsim{\mathrel{\rlap{\lower3pt\hbox{\hskip0pt$\sim$}}
    \raise1pt\hbox{$<$}}}         
\def\gsim{\mathrel{\rlap{\lower4pt\hbox{\hskip1pt$\sim$}}
    \raise1pt\hbox{$>$}}}         

\begin{document}

\begin{flushright}
NYU-TH/02/08/10 \\
TPI-MINN-02/24, UMN-TH-2107/02\\
\end{flushright}

\begin{center}

{\Large \bf Diluting Cosmological Constant
\\
\vskip 0.2cm

~via  Large Distance Modification of Gravity}\footnote{Based on talks given by the authors
at various recent conferences:
2002 Aspen Winter Conference on Particle Physics
{\em Current and Upcoming Discoveries}, February 2002, Aspen, Colorado;
{\em Continuous Advances in QCD 2002/Arkadyfest},
May 17-23, 2002, Minneapolis, MN, USA;
 ICTP Summer School on Astroparticle Physics and Cosmology,  June 17-July 5
2002, Trieste, Italy;  Payresq Physics Workshop,
June 22-30, Payresq, France; Third International Sakharov Conference,
June 24-29, Moscow, Russia; XVIIIth IAP Colloquium on the Nature of Dark Energy, July 2002, Paris,
France; PIMS Workshop on Brane World 
and Supersymmetry {\em Frontiers in Mathematical Physics}, UBC, July 22--August 2, 2002, Vancouver, Canada.
} 

\vskip 0.5cm {Gia Dvali$^a$, Gregory Gabadadze$^{b,}$\footnote{
Address after September 1, 2002: {\em Theory Division, CERN, CH-1211,
Geneva 23, Switzerland}}, and M. Shifman$^b$}

\vskip 0.5cm
{\it $^a$Department of Physics, New York University, New York, NY 10003\\
$^b$Theoretical Physics Institute, University of Minnesota, Minneapolis,
MN 55455}\\
\end{center}

\begin{center}
{\bf Abstract}
\end{center}

We review a solution (Ref. \cite {one}) of  the cosmological
constant problem in a brane-world model with infinite-volume
extra dimensions. The solution is based on a nonlinear  generally
covariant theory of a metastable graviton that leads to a
large-distance modification of gravity.

From the extra-dimensional standpoint  the
problem is solved due to the fact that the four-dimensional
vacuum energy curves mostly the  extra space.
The four-dimensional curvature is small, being inversely proportional to
a  positive power of the vacuum energy.
The effects of infinite-volume extra dimensions are seen by a
brane-world observer as nonlocal operators.

From the four-dimensional perspective the problem is solved
because the zero-mode graviton is extremely weakly coupled to localized
four-dimensional sources. The observable gravity is mediated
not by  zero mode but, instead, by a metastable graviton
with a lifetime of the order of the present-day Hubble
scale. Therefore, laws of gravity are modified in the infrared above
the Hubble scale. Large wave-length
sources, such as the vacuum energy, feel only the zero-mode
interaction and, as a result, curve   space
very mildly. Shorter wave-length sources interact predominantly via exchange
of the metastable graviton. Because of this, all   standard properties
of early cosmology, including inflation, are intact.

\vspace{0.1in}

\newpage

\section{Introduction}

The cosmological constant problem can be cast into   two
questions. First, there is an  old question:

 (i)  Why is the vacuum energy determined by a
momentum scale
  much smaller than any reasonable cut-off scale in
effective field theory of particle interactions?

This is sometimes referred to as the ``old" cosmological constant
problem. In   view of the present astrophysical
observations \cite {cc} one should  also
find  an  answer to the second question:

 (ii)  How come that the vacuum energy and the matter energy
are comparable today? Do we live in a special epoch?

This is the so-called ``cosmic coincidence" problem.
To our knowledge the only existing framework, which can address
  both questions simultaneously, is the  anthropic approach
\cite {WeinbergPRL,Vilenkin}.  Below we will
concentrate on a dynamical solution of the ``old'' cosmological
constant problem suggested in \cite {one}  which is based on a
nonlinear and generally covariant model of   metastable
graviton proposed in Refs. \cite {DGP,DG}.
We have nothing to say about the ``cosmic coincidence'' problem.

Before reviewing the solution of Ref. \cite {one}
 let us formulate the question properly.
As it stands, the question  (i)  is ill-posed
for our purposes. Let us first recall why the vacuum energy in
the universe   is normally assumed to be small.

In general relativity the cosmological expansion of the universe is
described by a standard metric
\beq
ds^2\, = \,dt^2\, -\, a^2(t)\, d{\vec x}^2\,.
\label{frw}
\eeq
Here $a(t)$ is the scale factor and $t$ is time in the
co-moving coordinate system (we assume for definiteness
that three-dimensional curvature is zero). In the
presence of the vacuum energy density $\E$, the scale factor
$a(t)$ obeys the Friedmann equation
\beq
H^2\, \equiv \, \left ( {{\dot a}\over a}\right )^2\, = \,
{\E \over 3\,\mpl^2}\,,
\label{Friedmannn}
\eeq
where ${\dot a}\equiv d\, a/dt$ (for simplicity we set $\E>0$).

We have no direct experimental way to measure $\E$. Instead,
we measure space-time curvature by cosmological observations, and
then determine $\E$ through Eq. (\ref{Friedmannn}).
Thus, claiming that $\E$ should be small we implicitly assume
that Eq. (\ref{Friedmannn}) is valid for arbitrarily large
length scales. This assumption does not need to be true.

Many approaches to the cosmological constant problem
were designed  to  dynamically cancel the right-hand side of
Eq. (\ref{Friedmannn}). Within   four-dimensional
local field theories with  a finite number of degrees of freedom
dynamical cancellation is ruled out by Weinberg's no-go theorem
\cite {Weinberg}.  We take an alternative route.
Following \cite {one}, we accept that the vacuum energy in our
world can be  large, but
we question the validity of Eq. (\ref {Friedmannn}) in the 
extreme low-energy approximation.

We use the framework of Refs.
\cite {DGP,DG} where gravity in general,
and the Friedmann equation, in particular,
is modified for   wave lengths larger than a certain critical value.
The cosmological constant problem is then remedied
in the following way: Due to large-distance modification of gravity
the energy density $\E\gsim (1 \,\, {\rm TeV})^4$  does not curve the space
as it would in the conventional Einstein gravity.
Therefore, the observed space-time curvature is
small, despite the fact that $\E$ is huge (as it comes out naturally).
This is the most crucial point of the  approach of Ref. \cite {one} ---
the point where we depart from the previous investigations.

Before we come to details,
let us briefly discuss the Weinberg no-go theorem \cite {Weinberg}
adapted to the present case (for details see Ref.~\cite {one}). The theorem states that
$\E$ cannot be canceled (without fine-tuning) in any
effective four-dimensional theory that satisfies the following conditions:

(a) General  covariance;

(b) Conventional four-dimensional gravity is mediated by a
{\it massless} graviton;

(c) Theory contains a finite number of fields below the cut-off scale;

(d) Theory contains no negative norm states.

Since we do not try to cancel $\E$  but rather intend to 
suppress the space-time
curvature induced by it, one  could have argued that
the theorem is inapplicable to begin with.
However, this argument is
incomplete and unsatisfactory. The point is that in any theory
obeying  conditions (a-d) Eq. (\ref{Friedmannn}) is  valid.
Therefore, in any such theory   small $H^2$ would require small $\E$.

We conclude that a successful solution must violate at least one
condition in (a-d). The solution of Ref. \cite {one}
does violate (b) and (c). In particular,
the zero-mode graviton, that mediates gravity in the far-infrared,
has a coupling which is many orders of magnitude smaller than
the Newton coupling. That is why   large value of $\E$
does not induce huge curvature.

The mode that mediates gravity   between shorter wave-length
sources, such as  most of the sources  in the observable
part of the universe, is coupled to the sources with the  usual
Newton constant. This gives rise to  conventional interactions
for observable sources in the universe.

Note that the division in two different modes is purely for convenience.
In general one could just drop the mode expansion language altogether
and state that the  strength of gravity depends on the wave length of
the source. Although in the present case   condition (c)
is violated, in other possible realizations of
the present idea this may not be necessary. At the same time,  
violation of condition (b) seems   inevitable in
any model that solves the cosmological constant problem
through large-distance modification of gravity.

In four-dimensional
general relativity  gravitational interactions are mediated
(at least at distances $r \gsim $ 0.1 mm)
by a massless spin-2 particle, the graviton $h_{\mu\nu}$.
General covariance (and the absence of
ghosts and tachyons) requires the universality of  the
graviton coupling to matter
\beq
  h_{\mu\nu}\,T^{\mu\nu}\,.
\eeq
General covariance also fixes  uniquely the low-energy
effective action   to be the Einstein-Hilbert action
\beq
{S}\,=\,{\mpl^2 \over 2}
\,\int\,d^4x\sqrt{g}\, \left ( R\,- 2\Lambda \right )\,,
\label{EH}
\eeq
where the cosmological constant $\Lambda = -\E/\mpl^2$
is included. The universal coupling to all sorts of energy,
including the vacuum energy, is the reason for the emergence of the
cosmological  constant problem.

If the measured four-dimensional
gravity were not mediated by an exactly
massless state, the universality of coupling could be avoided.
Thus, one may hope that in  such  theories   very large wave-length sources
(such as the vacuum energy) may effectively decouple from four-dimensional
gravity, eliminating the cosmological constant problem.
This is what happens in theories with infinite-volume
extra dimensions. This phenomenon can be understood from the
point of view of a four-dimensional brane observer as a modification of
the equation of motion for the graviton. In conventional linearized
general relativity the graviton obeys the following free field equation:
\beq
\bo \, h_{\mu\nu}\,=\,0 \,,
\label{free}
\eeq
while in the present case \cite {DGHS,one} (see also Sec. 4)
the graviton obeys a modified linearized equation
\beq
\left\{ 1 +
{\m^{N-2} \over c_1(N)\,\m^{N-2} + c_2(N)
\left(\bo \right )^{N-2\over 2}}\,{1 \over r_c^2\,\bo}
\right\} \,\bo \, h_{\mu\nu}\,= \,0\,,
\label{free1}
\eeq
where $\m$ is the scale of higher dimensional theory,
$c_{1,2}$ are some constants, and $N$ denotes the number of extra
dimensions.
$r_c = \mpl/ \m^2$ is the critical crossover distance beyond which
gravity is modified, as described by Eq.~(\ref{free1}).
The effective strength of gravity is set
by the operator in the   braces, which  tends to unity for the short wave length
($\ll \, r_c$) gravitons. As a result,
Eq.~(\ref{free1}) becomes indistinguishable from  Eq. (\ref{free}).
Thus, in this region, the theory
reproduces   predictions of  ordinary gravity with 
the standard $G_N$ coupling.
However, in the deep-infrared region, where the momenta are smaller than
$r_c^{-1}$,   new terms in (\ref {free1})
dominate over the conventional term  (\ref {free}).
Hence,  infrared physics is modified. It is this modification that
serves as a loophole in the no-go theorem for solving the cosmological
constant problem. It is remarkable that such a modification takes place
in a manifestly generally covariant theory.

\section{A brief overview of the model}

The effective low-energy action in these theories takes the form
\cite {DGP,DG}
\beq
{S}\,=\, \m^{2+N}\,\int\,d^4x\,d^N\rho\,\sqrt{G}\,{\cal R}\,+\,\int\, d^4x
\sqrt{\bar g}\,\left ({\cal E}\,+\,M^2_{\rm ind}\, R\,
+\,{\cal L}_{\rm SM}\right )\,.
\label{actD}
\eeq
Here  $G_{AB}$  stands for a
$(4+N)$-dimensional metric $(A,B=0,1,2,...,3+N)$, while $\rho$
are ``perpendicular'' coordinates.
For simplicity we do not consider brane
fluctuations. Thus, the induced metric on the brane is given by
\beq
{\bar g}_{\mu\nu}(x)~\equiv~G_{\mu\nu}(x, \rho_n=0), \qquad (n=4,...,3+N)\,.
\label{ind}
\eeq
The first term in (\ref{actD}) is the bulk Einstein-Hilbert  action
for $(4 +N)$-dimensional gravity, with  the fundamental scale $\m$.
The expression in  (\ref{actD}) has to be understood
as an effective low-energy Lagrangian valid for graviton momenta
smaller than $\m$. We imply that, in addition,
there are an infinite number of
gauge-invariant high-dimensional
bulk operators suppressed by inverse powers of $\m$.

The   $M^2_{\rm ind}\, R$ term in
(\ref{actD}) is the four-dimensional
Einstein-Hilbert (EH) term of the induced
metric. This term plays a crucial role.
It ensures that  the laws of four-dimensional gravity are reproduced
at observable distances
on the brane
despite the fact that there is no localized
zero-mode graviton.  Its coefficient $M_{\rm ind}$ is another
parameter of the model. $\E$ denotes the brane tension.
Thus,  the  low-energy action, as it
stands in (\ref {actD}), is governed by three parameters
$\m$, $M_{\rm ind}$ and ${\cal E}$. Both parameters $\m$ and $M_{\rm
ind}$ are perturbatively-stable under quantum corrections.
The parameter ${\cal E}$
plays the role of the vacuum energy.

 For ${\cal E} = 0$ the
gravitational dynamics on the brane
is as follows. The infinite extra space notwithstanding, a
brane observer   measures
four-dimensional gravitational interaction up to a certain
critical distance $r_c$. The
potential between   two static sources on the brane scales as
\beq
V(r)\, \propto \, -\, {1 \over M^2_{\rm ind}\, r}\,,
\label{Newton}
\eeq
for distances in the interval
\beq
\m^{-1}\,\lsim \, r \, \lsim \, r_c \,,
\eeq
where
\beq
 r_c\sim \left\{
\begin{array}{ll}
M_{\rm ind}^2/\m^3\quad \mbox{for}\quad N=1\\[0.2cm]
M_{\rm ind}/\m^2\quad \mbox{for}\quad N>1
\end{array}
\right. \,.
\eeq
However, at distances smaller than $\m^{-1}$ and
larger than $r_c$ gravity changes. An important requirement
is that gravity must be soft above the
scale $\m$ \cite {DGKN}.
The expression (\ref {Newton}) fixes $M^2_{\rm ind} = \mpl^2/2$.
In order for the late-time cosmology to be standard
we require  that
$r_c ~\sim ~H_0^{-1}\sim 10^{28}~{\rm cm}$.
This restricts the value of the bulk gravity scale to lie
in the ball-park
\beq
\begin{array}{ll}
10^{-3}~ {\rm eV}\,\lsim \,  \m \, \lsim \,
100 \,\, {\rm MeV}\quad\mbox{for}\quad N=1\,,\\[2mm]
 \m \sim 10^{-3} ~{\rm eV}\quad\mbox{for}\quad N>1\,.
\end{array}
\label{bounds}
\eeq
Inclusion of non-vanishing  ${\cal E} \gg \m^4$ triggers inflation
on the brane. The inflation rate  is rather peculiar and is
given by \cite {one}
\beq
H^2\,\sim \,\m^2\,\left (\m^4 \over \E   \right )^{2\over N-2}\,.
\label{Hinv}
\eeq
(This formula is not applicable to the $N=2$ case; see Ref. \cite {Zura}
for a discussion of induced gravity in this case.)
Therefore, a large vacuum energy density $\E$ causes
a small rate of inflation for $N> 2$.

The following question immediately come to one's mind:
How can such a behavior be understood from the standpoint
of a four-dimensional observer on the brane?
We will answer this question using the
linearized analysis of \cite{DGP}.

The effect can be best understood in terms of the four-dimensional
mode expansion. From this perspective a high-dimensional
graviton represents a continuum of four-dimensional states and
can be expanded in these states. Below we will be interested
only in spin-2 components for which the KK decomposition  can
schematically
be written as follows:
\begin{equation}
  h_{\mu\nu}(x,\rho_n)\, = \,\int d^N m\, \epsilon_{\mu\nu}^m(x)\,
 \sigma_m (\rho_n)\,,
\label{kkexp}
\end{equation}
where  $\epsilon_{\mu\nu}^m(x)$ are four-dimensional spin-2 fields of mass
$m$ and   $\sigma_m(\rho_n)$ are their wave-function profiles
in extra dimensions. The strength of    individual mode coupling  
to a brane source is given by the value of the wave function at the
position of the brane, that is  $\sigma_m(0)$.
 Four-dimensional gravity on the brane is mediated by
exchange of all the above modes. Each of these modes gives rise to
a Yukawa type gravitational  potential. The net result is
\begin{equation}
 V(r)\,\propto \,{1 \over \m^{2 + N}}\,
\int_0^\infty dm\, m^{N-1}\, |\sigma_m(0)|^2
\,{e^{-rm} \over r}\,.
\label{kkpot}
\end{equation}
It is a crucial property of the model (\ref {actD}) that
four-dimensional gravity on the brane is recovered
for $r\ll r_c$ due to the fact that   modes with
$m > 1/r_c$ have suppressed wave-functions \cite {Wagner,DGKN1}
and, therefore, the  above integral is effectively cut-off
at the upper limit at $m \sim 1/r_c$.
Most easily this can be seen from the propagator analysis.
Gravitational potential (\ref{kkpot}) on the brane is mediated by an
``effective''   4D graviton which can be defined as
\begin{equation}
h_{\mu\nu}(x,0)\, = \,\int d^N m\, \epsilon_{\mu\nu}^m(x)\, \sigma_m
(0)\,.
\label{effgraviton}
\end{equation}
The Green's function for this state can be defined in a usual way.
Using (\ref{effgraviton}) and orthogonality of the
$\epsilon_{\mu\nu}^m(x)$ states we obtain
\begin{equation}
{\cal G}(x-x',0)_{\mu\nu,\gamma\delta}\, = \,
\langle  h_{\mu\nu}(x,0) \,h_{\gamma\delta}(x',0) \rangle =
 \int d^N m |\sigma_m(0)|^2
 \langle\epsilon_{\mu\nu}^m(x) \epsilon_{\gamma\delta}^m(x')\rangle \,.
\label{propeffective}
\end{equation}
From now on we will suppress the tensor structure, which is not
essential for this discussion. Passing to the Euclidean
momentum space we get the following expression for the scalar part
of the propagator
\begin{equation}
G(p,0) \, = \,\int dm \, m^{N-1} \,{|\sigma_m (0)|^2 \over m^2 + p^2}\,.
\label{propeffm}
\end{equation}
This is the spectral representation for the
Green's function
\begin{equation}
G(p,0)\,  = \,\int d s \,{\rho(s) \over s + p^2}\,,
\label{KLR}
\end{equation}
with $s \equiv m^2$ and
\begin{equation}
\rho(s) \, = \,{1\over 2}\, s^{{N-2\over 2}} |\sigma_{\sqrt{s}} (0)|^2\,.
\label{rhosigma}
\end{equation}
Therefore, the spectral  representation
can be simply understood as the Kaluza-Klein mode expansion
(\ref{kkexp}). Then, the wave-function suppression of the heavy modes
can be read off from Eqs. (\ref{propeffm}) and (\ref{KLR}) by using the
explicit form of the propagator $G(p)$ \cite{DG,DGHS},
\beq
G(p, 0 )\,=\,{1 \over \mpl^2 p^2\,+\,\m^{2+N} D^{-1}(p, 0)}\,,
\label{gee111}
\eeq
where $D^{-1}(p, 0)$ is the inverse Green's function
of the bulk theory with no brane.
For the purposes of the present discussion it is enough to note
that at large momenta $p \gg r_c^{-1}$
the above propagator behaves as \cite {DGP,DG}
\beq
G(p, 0 )\,\simeq\,{1 \over \mpl^2 \,p^2},
\label{four-dimensionalp}
\eeq
which is the propagator of a
massless four-dimensional graviton with
the  coupling $1/\mpl^2$. Substituting (\ref{four-dimensionalp}) in  the left-hand side
of (\ref{KLR}) we find  that the function $\rho(s)$ must be
suppressed  at $s\gg  r_c^{-2}$. If so, the relation (\ref{rhosigma})
implies that the wave functions of the heavy modes must be
vanishingly small as well.

For the $N=1$ case
both the propagator \cite{DGP} and the wave-function profiles
can be evaluated analytically \cite{DGKN1},
\beq
G(p, 0 )\,=\,{1 \over \mpl^2\, p^2\, + \,2\m^3\, p}\,,
\label{N1prop}
\eeq
and
\beq
|\sigma_m(0)|^2\, = \,{4 \over 4 + m^2{\mpl^4 /\m^6}}\,.
\label{N2prof}
\eeq
This shows  that all the modes which  are
heavier than $r^{-1}_c = {\m^3/\mpl^2}$
are suppressed on the brane.
Substituting (\ref{N2prof}) into  (\ref{kkpot}) we derive the
usual Newton  potential (\ref {Newton}) at distances $r \ll  r_c$.

One can interpret the above Green's function
as describing  a metastable state that decays into the bulk states
with the lifetime $\tau_c \sim r_c$. A remarkable fact is that the
existence of such  a metastable state
is perfectly compatible with the  exact four-dimensional
general covariance.

\section{Dilution of the cosmological constant}

It is instructive to rederive the above-mentioned results
with extra dimensions being compactified at very large
cosmological distances.
For  non-vanishing $\E$ the compactification is not trivial.
If we were to set the compactification radius $L$
smaller than the gravitational radius of the brane
\cite {Gregory,Emparan}
\beq
\rho_g\,\sim \, \m^{-1}\,\left ({\E \over \m^4} \right )^{1\over N-2}\,,
\label{rhog}
\eeq
this would distort the brane gravitational
background very strongly. Therefore,
$L$ has to be at least somewhat larger
than $\rho_g$. For  realistic values of $\E$ and $\m$
this leads to an estimate
$L\gsim H_0^{-1}\sim 10^{28} \,\, {\rm cm}$ \cite {one}.
In what follows for simplicity we will not distinguish
the values of $L$, $\rho_g$ and $H_0^{-1}$, in spite of the fact
that there should be an order of magnitude
difference between these scales ($L\gsim \rho_g \gsim H_0^{-1}$).
We simply set all these values in
the ball-park of $H_0^{-1}$. Let us now address the question: what does
a 4D observer see on the brane?

Because the space is compactified, there is a 
mass gap in the KK modes. Start from the
zero-mode massless
graviton. This mode couples universally to the brane matter
and vacuum energy. However, because the compactification scale is
huge $L\sim \rho_g\sim H_0^{-1}$ the coupling of the zero mode
$G_{\rm zm}$ is tiny. Indeed,
\beq
G_{\rm zm}\,\sim \, {1 \over \m^{2+N}\,L^N\,+\,\mpl^2}\,.
\label{Gzm}
\eeq
The first term in the denominator is due to the
compactness of the extra space and the second term is
due to the induced EH term in (\ref {actD}).
In particular, $\m^{2+N}\,L^N \gg \mpl^2 $ and, therefore,
\beq
G_{\rm zm}\,\ll \,G_N\,.
\label{zmN}
\eeq

Besides the zero mode, there is a tower of massive KK modes,
with masses quantized in the units of $H_0$.
{\it A priori} each of these modes is important for interactions
at  observable distances $\ll L$, for instance in the solar system.
However, due to the presence of the induced EH term on
the brane (\ref {actD}) the wave functions of these
modes are suppressed on the 4D world volume. The net result
due to the KK modes can now be summarized as a single
metastable graviton with the lifetime $\sim L\sim H_0^{-1}$.
The two descriptions ---  one in terms of the KK modes, and the other one
in terms of the metastable  graviton --- are complimentary to each other.

The coupling of this metastable graviton to matter is determined
by the coefficient in front of the induced EH term in
(\ref {actD}). Since the latter is set to be $\mpl$,
the metastable mode couples to matter with the
Newton coupling $G_N$. Therefore, at observable distances,
that are somewhat smaller than $10^{28}\,\,{\rm cm}$,
the laws of 4D gravity are enforced by a metastable graviton
which at these scales can be treated as a (almost) stable
spin-2  state interacting with $G_N$. For these scales
the presence of the zero mode is irrelevant
because of its tiny coupling $G_{\rm zm}$. This
provides  conventional gravity and cosmology for any
time scale up until today.

Let us now turn to  distances  somewhat
larger that $10^{28}\,{\rm cm}$. There,
the metastable mode does not mediate interactions
(since its lifetime is less than $H_0^{-1}$).
Instead, one should think in terms of the massive KK modes.
Since the mass of the lightest KK mode is  $\sim H_0$,
the interactions due to the KK modes are exponentially suppressed
at distances larger than  $H_0^{-1}$.
Therefore, one is left with
the zero mode interactions only. Any source which has
characteristic wave lengths larger than $H_0^{-1}$,
(i.e., the characteristic momenta
smaller than $H_0$) will feel gravity only due to
the zero mode. Because the  coupling of the zero mode is tiny (\ref {Gzm}),
the curvature $R$ produced by this source will also be small
\beq
R\, \sim \,{G_{\rm zm}\,\E}\,\sim \,{\E \over \m^{2+N}\,L^N}\,\ll
{\E \over \mpl^2}\,.
\label{R}
\eeq
Since  $R\sim H^2$ and $L\sim \rho_g$
where $\rho_g$ is defined in (\ref {rhog}) we find from (\ref {R})
\beq
H^2\,\sim \,\m^2\,\left (\m^4 \over \E   \right )^{2\over N-2}\,.
\label{Hcom}
\eeq
This coincides with (\ref {Hinv}).
Therefore, the small inflation rate (small  curvature) is
due to the fact that  the zero mode is very weakly coupled
to vacuum energy.  Summarizing, compactification of extra
dimensions with the ultra-large radius (\ref{rhog})
produces essentially the same physical picture as in the model with
infinite extra dimensions.

\section{Four-dimensional picture on the brane}

In this section we return  to the uncompactified case.
Four-dimensional gravitational interactions
are mediated by a gapless infinite tower
of Kaluza-Klein modes the wave functions of
which are suppressed on the brane.
Alternatively, these interactions can be
thought of to be mediated by a {\it single metastable}
four-dimensional graviton.
The expression for  the two-point Green's function
on the brane \cite {DGP} leads to  the following
equation for this four-dimensional ``effective'' graviton
\cite {DDG}:
\beq
\left (
\bo\,+\,m_c\,\sqrt{\bo }\,\right )\,h_{\mu\nu}\,=\,
\p\,\left (
T_{\mu\nu}\,
-\,{1\over 3}\,\eta_{\mu\nu}\,T\right )\,.
\label{pert0}
\eeq
Here $T\equiv T_{\nu}^{\nu}$ and $m_c\,\equiv \,r_c^{-1}\,\sim\,\m^3/\mpl^2$.
This refers to one extra dimension\footnote{Two branches of the square root
in (\ref {pert0}) lead  to the standard and self-inflationary cosmological
solutions \cite {Deffayet}.}, $N=1$. As before,
$\m$ denotes an ultraviolet scale at which gravity
breaks down  as an effective
field theory.  On phenomenological grounds,
this can be any scale in the interval $ 10^{-3}\,{\rm eV}\,\lsim
\m\,\lsim \,\mpl$. As we discussed before, in the
five-dimensional model we have $ 10^{-3}\,{\rm eV}\,\lsim
\m\,\lsim \,100\,\,{\rm MeV}$, so that
$m_c \lsim H_0\sim 10^{-42}\,\,{\rm GeV}$
is less than the Hubble scale $H_0$.

We would like to provide a nonlinear completion of Eq. (\ref {pert0}).
Let us start with the right-hand side (r.h.s.) of the equation.
It is known that in the massive (metastable) graviton theory  
the tensorial structure of the graviton propagator
is affected by {\it nonlinear} corrections \cite {Arkady, DDGV}.
The expression  which takes  these nonlinear effects into account
can be parametrized as follows:
\beq
{\rm r.h.s.}\,=\,\p\,\left\{ T_{\mu\nu}\,
-\,{\cal K}\left(\frac{m_c}{\mu}\right)\eta_{\mu\nu}\,T \right\}\,,
\label{pertk}
\eeq
where $ {\cal K}(m_c/\mu) $ is a function of the physical scale
$\mu$ set by the source $T$ and of the distance
from the source \cite {Arkady,DDGV}.
For large (unobservable) distances it gives rise to
$ {\cal K}(m_c/\mu) \simeq 1/3$, i.e.,
the result in (\ref {pert0}). However,
at measurable distances (e.g. solar system)
we have $ {\cal K}(m_c/\mu) \simeq 1/2$ due to nonlinear effects.
Hence, the discontinuity in the mass parameter $m_c$
\cite {Iwasaki,Veltman,Zakharov}
is in fact absent in   nonlinear theory of massive gravity
\cite {Arkady} and, in particular, in the model of metastable
graviton \cite {DGP},
as was shown in \cite {DDGV,Lue,Gruzinov,Porrati}
(see also \cite{Ian}).  Therefore,
for the distances of practical interest
\beq
{\rm r.h.s.}\,=\,\p\,\left ( T_{\mu\nu}\,
-\,{1\over 2}\,\eta_{\mu\nu}\,T\right )\,,
\label{pert2}
\eeq
which is a correct tensorial structure of the conventional
Einstein equation.

The next step is to perform a nonlinear completion
of the left-hand side (l.h.s.) of Eq. (\ref {pertk}).
The latter procedure is somewhat arbitrary
from the point of view of pure 4D theory, however
it is uniquely fixed by the higher dimensional theory
\cite {Porrati}. Here we are  interested in qualitative features
which are independent of the form of the
completion. To this end one can use the substitution
\beq
h_{\mu\nu} \,\to\,{1\over \bo}\,R_{\mu\nu}\,,
\label{hR}
\eeq
where $R_{\mu\nu}$ denotes the Ricci tensor.
Furthermore, performing simple
algebra we find the following nonlinear
completion of the Eq. (\ref {pertk}):
\beq
{\cal G}_{\mu\nu}\,+\,{m_c\over \sqrt{\bo}}\,{\cal G}_{\mu\nu}\,=\,
\,\p\,T_{\mu\nu}\,,
\label{nlGfive-dimensional}
\eeq
where ${\cal G}_{\mu\nu} \equiv R_{\mu\nu} -(1/2) g_{\mu\nu}R$.
This should be compared with the conventional Einstein equation
\beq
{\cal G}_{\mu\nu}\,=\,\p\,T_{\mu\nu}\,.
\label{einst}
\eeq
Note that the second term on the l.h.s in Eq. (\ref {nlGfive-dimensional})
does not appear in the Einstein equation.
In fact, this is the very same term that dominates in the infrared region.
The Bianchi identity imposes a new constraint on possible
gravitational backgrounds $m_c\nabla^\mu (\bo)^{-1/2}{\cal G}_{\mu\nu}=0$
(in the linearized approximation the latter  reduces to a gauge
fixing condition). One can also note that from
the standpoint  of four-dimensional theory  
  a violation of unitarity takes place. This corresponds to the fact that
the metastable graviton can decay into KK modes.
It is clear that the full five-dimensional unitarity is
preserved while it can be violated in any given four-dimensional
subspace.

As we have already mentioned,  
 the procedure of the nonlinear completion
used above is not unique. From the 4D point of view there is no guiding principle
  for this completion.
In general, an infinite number of nonlinear terms
with arbitrary coefficients  can be added to  the r.h.s. of
(\ref {hR}). This would lead to an infinite number of new terms on the
l.h.s. of Eq. (\ref {nlGfive-dimensional}).
However, what is critical is that there is a {\it unique} set of these terms
which complete the 4D theory to a higher dimensional theory
that we started from. The purpose of the exercise
performed above is to demonstrate that the presence of these
nonlocal terms leads to modification of gravity in the infrared.

Similar arguments apply to  higher co-dimensions.
In a model with $(4+N)$   dimensions \cite {DG}
the equation   analogous to   (\ref {pertk}) takes the form \cite {one}
\beq
\left\{ \bo +
{\m^{N+2}/\mpl^2 \over c_1(N)\,\m^{N-2} + c_2(N)
\left(\bo \right )^{N-2\over 2}}
\right\} \,h_{\mu\nu}=\p\left\{T_{\mu\nu}\,
- {\cal K}\left(\frac{m_c}{\mu}\right)\eta_{\mu\nu}\,T\right\},
\label{pertN}
\eeq
where $c_{1,2}(N)$ are some constants that depend on
$N$ \cite {Wagner,DGHS} (we do not discuss here the $N=2$
case for which there are logarithmic functions of $\bo$ on the l.h.s.
in (\ref {pertN})).
Using the same method as above we propose a nonlinear completion
of the equation which looks as follows:
\beq
{\cal G}_{\mu\nu}\,+\,
\left \{ {\m^{N+2}/\mpl^2 \over c_1(N)\,\m^{N-2}\,+\,c_2(N)\,
\left(\bo \right )^{N-2\over 2}}\right \}\,
{1\over {\bo}}\,{\cal G}_{\mu\nu}\,=\,\p
\,T_{\mu\nu}\,.
\label{nlGN}
\eeq
One  observes the same pattern: the second term on the l.h.s.
dominates in the infrared. Hence, infrared gravity is modified.
This is a necessary condition for the present approach
to solve the cosmological constant problem. However,
it is not sufficient,   generally speaking. As we argued in Ref. \cite {one},
and in the previous sections, only $N>2$ models can solve
the cosmological constant problem.

\section{Killing cosmological constant does not kill inflation}

The present scenario solves the cosmological constant problem due to
modification of 4D gravity at very low energies. Due to
this modification
a strictly constant vacuum energy curves four-dimensional space
extremely weakly. As we have seen, the contribution to the effective
Hubble expansion rate from the vacuum energy ${\cal E}$ is set by
an inverse power of ${\cal E}$, see Eq. (\ref {Hinv}).
A question  to be discussed in this section
is whether the dilution of cosmological constant kills
inflation. Naively this seems inevitable. According to the
inflationary scenario, our 4D universe underwent the period of
an exponentially fast expansion (for a review see \cite{linde}).
In the standard case this is
achieved by introducing a slow-rolling scalar field, the inflaton.
During the slow-roll epoch the potential energy dominates
and mimics an approximately constant vacuum energy. This leads to
an accelerated growth of
the scale factor.

However, as we have just argued, in the present case any
constant vacuum energy is diluted according to
Eq. (\ref {Hinv}). If so,  it seems that the same
should happen to any inflationary energy density. In other words, one
would naively expect that standard inflation with the
energy density ${\cal E}_{\rm inf}$ should
generate the acceleration rate
$ H\,\sim\,\m\,\left (  {\m^4 / {\cal E}_{\rm inf}  } \right
)^{1/ N-2}\,$.
If such a relation indeed took place during the slow-roll period
this would eliminate inflation and all the benefits one gets with it.
Fortunately, this is not the case as we will now argue.

The relation (\ref {Hinv}) is only true
for energy sources with wave lengths $\gg r_c \sim H_0^{-1}$.
Brane sources of shorter wave lengths gravitate
according to the conventional 4D laws since they are coupled to
the resonance graviton. Furthermore,
let us recall that ${\cal E}_{\rm inf}$  is {\it time-dependent}
during slow-roll
inflation, with a typical time scale
that is  smaller than $r_c$ by many orders of magnitude.
For instance, consider
inflation with ${\cal E}_{\rm inf} \gsim (1 \,\, {\rm TeV})^4$.
For such
inflation to solve the horizon and flatness problems and generate
density perturbations, ${\cal E}_{\rm inf}(t)$ must be approximately
constant on the time scale of the inflationary Hubble time $\sim
M_P/\sqrt{{\cal E}_{\rm inf}}\, \lsim  \, {\rm mm}$. This
is at least $30$ orders of magnitude smaller than our $r_c$.
That is why inflation in our scenario
  proceeds  in the same manner as in the conventional setup.
Below we will consider this issue in more detail.

Let us first consider the simplest nonlinearly-completed effective 4D
equation (\ref {nlGN}). Although  it is
equivalent to the high-dimensional model (\ref {actD})
only in the linearized approximation, nevertheless, it
captures the most essential feature ---
the large-distance modification of gravity.
Taking the trace of the above equation
we get the following relation between the curvature and
stress tensor
\begin{equation}
\left[1 \,+ \,{M_*^{N-2} \over c_1\,M_*^{N-2} + c_2\,\nabla^{N-2}}
\,{1 \over (r_c\,\nabla)^2} \right]\, R \,= \,8\pi G_N\, T_{\mu}^{\mu}\,.
\label{RT}
\end{equation}
Let us consider a constant vacuum energy
${\cal E}$ as the source in the r.h.s. of this equation.
In the conventional Einstein gravity
this would induce curvature $R \sim  G_N{\cal E}$.

However,  this cannot be a solution in our case since the second term
on the l.h.s. diverges for a constant curvature.
This divergence  is just an artifact of our simplified
non-linear completion in which we keep only finite number of the
${\nabla}^{-1}$ operators. Careful resummation of all such
terms should give rise to a small but finite curvature as it
is apparent from
the high-dimensional description (see Eq. (\ref {Hinv})).

What is important, however, is the fact that the nonlocal
terms   dominate for constant curvature, while they are
negligible for time-dependent sources with characteristic
wavelengths $\ll r_c$. For instance,
let us assume $T_{\mu}^{\mu}$ is the energy density of a
slowly-rolling inflaton field. For the standard inflation, a
typical time scale of change in the energy density is by many orders of
magnitude smaller than any reasonable value of $r_c$. For
a  typical slow-roll inflation with $H \sim 10^{12}~ {\rm GeV}$
we get
\begin{equation}
\left[{1 \over (r_c\,\nabla)^2} \right]\, R \, \sim  \, 10^{-108}\, R\,.
\end{equation}
From this estimate it is clear that the nonlocal terms are
negligible, and the standard relation
is intact.

Were we able to calculate all   nonlocal terms in 4D
world volume, we would recover the series of the form
\begin{equation}
\left[1 + {M_*^{N-2} \over c_1\,M_*^{N-2} \,+\, c_2\,\nabla^{N-2}}
\,{1 \over (r_c\,\nabla)^2}
  \, + \, \sum_n {a_n \over (r_c\,\nabla)^n} \right]
\, R \, = \, 8\pi G_N \,T_{\mu}^{\mu} \,,
\label{RTfull}
\end{equation}
where the coefficients $a_n$ are, generally speaking,  functions of metric
invariants  that die away in the linearized flat-space limit.
For the time independent curvature, such as the one
induced by a constant vacuum
energy, each term diverges. Hence, explicit summation
must be performed   in (\ref {RTfull}).
In each particular case the summation should be possible, in principle.
However, in practice it may be much easier to solve directly the
high-dimensional equations. This, as we know, indicates that
the resulting curvature is given by Eq. (\ref {Hinv}) and is tiny.

The proportionality
of the curvature to the negative power of vacuum energy is
a peculiar property of large wave-length sources.
Nothing of the kind happens for   shorter wave-length sources.
For the slow-roll inflationary curvature the contribution
from nonlocal terms is completely negligible. This is why we expect that
inflation, as well as all   ``short distance" physics,
obey the conventional laws. A simple dimensional analysis shows
that   deviations  from the conventional behavior
must be suppressed by powers of $\lambda/r_c$, where
$\lambda$ is a typical wave length (or time scale) of the system.
Thus, our solution predicts measurable deviations from  predictions
of the Einstein gravity  for sufficiently large objects.
In fact this was explicitly demonstrated in Ref.
\cite {Gruzinov} in the  five-dimensional example
where the deviation for the Jupiter orbit might  be  potentially
observable.

\vspace{0.5cm}

{\bf Acknowledgments}
\vspace{0.1cm} \\

We would like to thank S. Dimopoulos,
A. Vainshtein and A. Vilenkin for useful discussions on the
subject of the paper. The work of G.D. is supported in
part by a David and Lucile  Packard Foundation Fellowship
for  Science and Engineering,
by Alfred P. Sloan foundation fellowship and by NSF grant
PHY-0070787. G.G. and M.S.
are supported by  DOE grant DE-FG02-94ER408.

\end{document}